# Negative differential electrical resistance of a rotational organic nanomotor

Hatef Sadeghi[*], Sara Sangtarash, Qusiy Al-Galiby, Rachel Sparks, Steven Bailey[*] and Colin J. Lambert

*Quantum Technology Center, Lancaster University, LA1 4YB Lancaster, UK*

* h.sadeghi@lancaster.ac.uk; s.bailey@lancaster.ac.uk

*Abstract-* A robust nanoelectromechanical switch is proposed based upon an asymmetric pendant moiety anchored to an organic backbone between two $C_{60}$ fullerenes, which in turn are connected to gold electrodes. Ab initio density functional calculations are used to demonstrate that an electric field induces rotation of the pendant group, leading to a non-linear current-voltage relation. The non-linearity is strong enough to lead to negative differential resistance at modest source-drain voltages.

**Introduction**

Biomotors utilising myosins, kinesins, and dyneins [1-4] have been utilised in several motor-protein driven devices for cargo transportation [5, 6] molecule sorting [7, 8], imaging [9] and sensing [10, 11]. In contrast to biological machines, which convert energy into directed motion by moving out of thermodynamic equilibrium [12-15], artificially designed nanoelectromechanical (NEM) motors operate by moving towards thermodynamic equilibrium. Many examples of artificial NEM devices use directed motion [16-26]. For example oscillators with frequencies in excess of 1 GHz have been constructed from multiwalled carbon nanotubes (MWCNT), where the telescoping nature of the inner carbon nanotubes [27, 28] with very low inter-wall friction [29-34] lead to novel electrical properties [35-40]. These examples illustrate how an electric field can induce motion and also how a motion-induced change of geometry can affect electrical properties. In what follows, our aim is to demonstrate that this coupling between a controlled geometry and electrical properties can lead to desirable non-linear current-voltage relations and negative differential resistance (NDR).

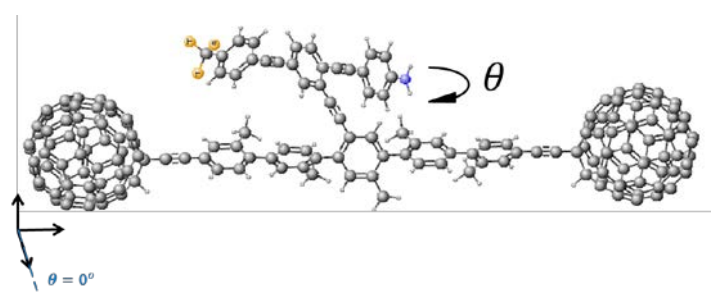

Figure 1 schematic of the proposed molecular switch where the asymmetric rotor blade is terminated at one end with nitrogen and at the other with three fluorine atoms with a single bond linking the rotor to the aromatic backbone. The two $C_{60}$ molecules act as secure anchors for the device and could be connected to gold leads.

As a specific example which demonstrates the general principle, we analyze the molecular-scale NEM shown in Fig. 1 whose conformation can be manipulated using an external electric field and whose conformation changes feedback to produce a non-linear current-voltage relation. This novel NEM consists of a pendant rotor attached by a single carbon bond to an aromatic backbone. The rotor is designed to possess a dipole moment aligned along its length such that an applied electric field will cause the rotor to turn relative to the aromatic backbone. Our aim is to examine the response of the device to external electric fields and determine the change in electrical conductance due to the associated conformational changes when the $C_{60}$s are attached to metallic electrodes. Our calculations will demonstrate that such conformation changes lead to negative differential resistance (NDR).

**Results and Discussion**

The dumbbell molecular switch shown in Fig.1 consists of three main sections, the backbone, the terminating groups and the branch. The backbone consists of five interconnected phenyl rings with attached methyl groups to prevent the backbone from twisting and is stabilized at either end by a fullerene, $C_{60}$ terminating group. The $C_{60}$ at either end not only stabilizes the molecule, but also allows the molecule to appear more clearly visible on STM images [41], therefore facilitating experimental STM measurements. The branch extends from the central ring of the backbone and is made up of three interlinked arene compounds; the central phenyl ring is capped by aniline at one end and terminated with a fluoro-toluene derivative at the other, where the hydrogens are replaced by fluorine. As fluorine is the most electronegative element, this design will enable the branch to possess a dipole moment. The dipole moment for the combined pendant group and backbone as an average of all the rotation angles of the pendant group relative to the backbone is approximately 9.4 Debye over a length of 28.23 Å. The length of the pendant group alone is 20.4 Å. The variation in the dipole moment, over all rotation angles, is given in figure SI-4 and the lengths in figure SI-5.

This dipole moment of the branch facilitates the electric field induced rotation required to create a switch. By applying an external electric field $E_{ext}$ across the molecule, the additional contribution to the total energy is $U = -p.E_{ext}$ where $p$ is the dipole and $E_{ext}$ is the external electric field. In the presence of a uniform electric field, the energy landscape of the system will change, with the possibility that the most stable rotation angle switches from one value to another. By computing the total energy as a function of rotation angle, we thereby obtain an estimate of the size of the electric field required to switch the molecule.

We use the *SIESTA* [42] implementation of the density functional theory (DFT) with a van der Waals density functional [43, 44] and extended and corrected double zeta polarised basis sets of the pseudo atomic orbitals. Geometries were optimised by relaxing the atomic forces to less than 20meV/Å. The van der Waals density functional allows long-range interactions to be taken into account. The total ground-state energy of the molecule is calculated to find the energy profile of the molecule with different confirmations. A basis set superposition correction is carried out to account for overlapping basis functions. This correction is calculated by taking the relaxed energy of the entire molecule and subtracting the energy of the structural relaxation of backbone and the branch separately: $U_{BSC} = U_{molecule} - U_{branch} - U_{backbone}$. Fig.2 shows the potential energy profile of the dumbbell molecule against rotation angle of the rotor with respect to the backbone. The DFT calculated energy profile yields an energy barrier to rotation of about *800 meV*.

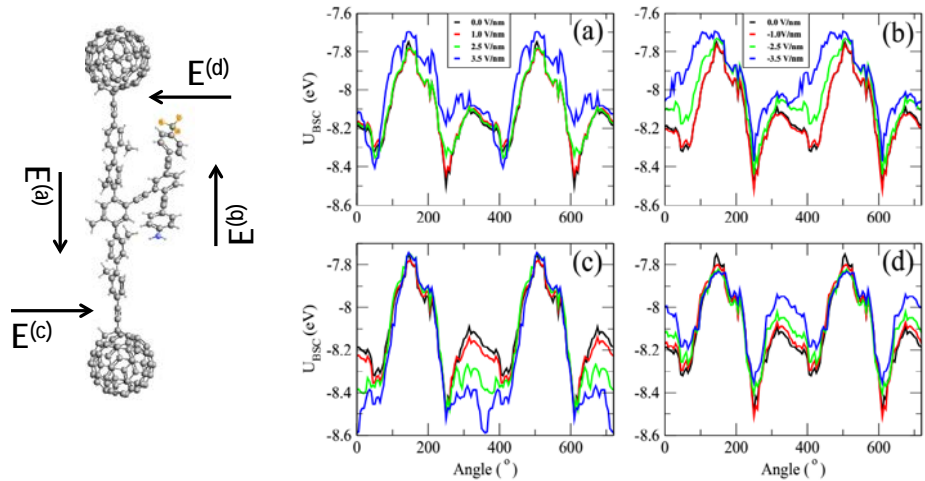

Figure 2 The potential energy profile $U_{BSC}$ (eV) calculated from the changes in the total energy of the system against rotation angle of the rotor with respect to the backbone as shown in Fig.1 for four applied magnitudes of electric field between 1.0 and 3.5 V/nm. Figures a, b, c and d correspond to the directions $E^{(a)}$, $E^{(b)}$, $E^{(c)}$ and $E^{(d)}$ shown in the left-most figure. The rotation angle is defined in figure 1 and figure SI-2. 240° corresponds to the pendant group being parallel to the backbone as shown in figure 2, whereas 60° (or equivalently 420°) corresponds to the pendant group making a 30° angle with the backbone. The zero-field energy minima are located at rotation angles of 60° and 240°. To sample all angles and compute these energy-versus-angle curves, the pendant group was rotated artificially to a chosen angle and then the molecule was allowed to relax to a local energy minimum.

Figure 2 shows the energy landscape as a function of rotation angle for four magnitudes of electric field between 1.0 and 3.5 V/nm. Figures 2a, 2b, 2c and 2d correspond to the directions $E^{(a)}$, $E^{(b)}$, $E^{(c)}$ and $E^{(d)}$ shown in the left-most figure. Only 2a and 2b are relevant to the two-terminal device shown in figure 3a. The plot in figure 2c and 2d are relevant to a three terminal device containing a gate electrode able to create a field perpendicular to the length of the molecule. In what follows, we focus on a two-terminal device only, since this is likely to be realised in the laboratory. At zero temperature, the rotation angle coincides with the global minimum of the energy curves. At finite temperature, the above minima correspond to the most-probable rotation angles, but other angles can be sampled, according to the Boltzmann factor (see equation 4 below). At zero field, the energy minima occur at $\theta_A = 60°$ and $\theta_B = 240°$. In both positions, the branch is parallel to the backbone and there is a significant overlap between one of the phenyl rings on the backbone with the aniline capped end of the branch. This suggests that these positions are stabilised by $\pi - \pi$ interaction between the two aromatic rings. The rotor does not rotate at a uniform distance from the backbone and therefore the charge distribution of the branch interacts non-uniformly with the backbone. This effect is apparent at $\theta_B$ where one end of the branch is located closer to the backbone than the other.

When an electric field is applied parallel to the backbone in the direction of $E^{(a)}$ (Figs. 2a), the global minimum at $\theta_B$ becomes a local minimum and the global minimum is located at $\theta_A$. Meaning that through applying an electric field parallel to the backbone, the most stable state of the molecule can be manipulated and the branch of the molecule will switch due to this electric field. One can also observe that by removing or reversing the direction of the electric field the branch can be switched back to $\theta_B$. As shown in Figs. 2c by applying an external field $E^{(c)}$ orthogonal to the molecule the global minimum of the energy curve can be switched to $0°$, whereas a field $E^{(d)}$ causes no such crossover.

To study the effect of the external electric field in transport properties of the dumbbell molecule, consider the molecule connected to two gold electrodes in a junction, as shown in Fig.3a. Since we are interested in two-terminal switch, we focus on the effect of the source-drain electric field induced by gold electrodes (electric field parallel to the molecule Fig.1a,b). By contacting the left and right $C_{60}$ to gold electrodes, electrons entering the leads from external reservoirs have Fermi distributions given by $f_L(E)$ and $f_R(E)$ and the Landauer formula [45] gives:

$$I = \frac{2e}{h} \int_{-\infty}^{+\infty} dE\, T(E) \left( f_L(E) - f_R(E) \right) \tag{1}$$

where the electronic charge $e = |e|$, $h$ is Plank's constant and $T(E)$ is the transmission coefficient for electrons with energy $E$ passing through the molecule from left ($L$) to right ($R$).

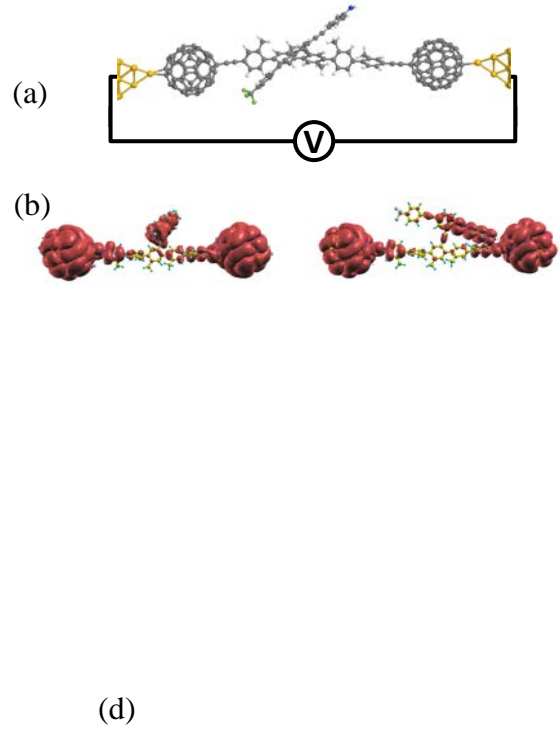

Figure 3. (a) the molecular structure within the junction. (b) A contour plot of the local density of states of the gas-phase molecule at $\theta=60^o$ (right) and $\theta=0^o$ (left). (c) and (d) A contour plot of the conductance $G/G_0$ and current $I/I_0$ ($I_0 = 2e^2/h \times 1 volt = 77 nA$) through the dumbbell molecule between two gold electrodes against angle $\theta^o$ and $E_F$ (eV) at room temperature.

The transmission coefficients $T(E)$ were calculated using *GOLLUM* [46] which is a newly developed simulation tool for electron, thermal and spin transport using the same method as described in [47]. Close to equilibrium $f_{L,R}(E) = (1 + exp(\mu_{L,R}))^{-1}$ where $\mu_{L,R} = \frac{E-E_F^{L,R}}{k_B T}$, $E_F^L$ ($E_F^R$) is the Fermi energy of the left (right) reservoir and $T$ is the temperature. As shown by the transmission curves in figure SI-1, transport is HOMO dominated. Figure 3b shows contour plots of the local density of states (LDOS) around the HOMO of the isolated molecule for two different rotation angles. These plots show that the HOMO is extended, but not symmetric. This demonstrates why the transmission coefficient does not approach unity on resonance, because it is well known that the transmission coefficient is less than unity in asymmetric systems, such as the structure in Figure 1 (see eg [47-49]). At zero temperature and finite voltage $E_F^L = E_F + \frac{eV}{2}$ and $E_F^R = E_F - \frac{eV}{2}$ the current could be calculated as,

$$I = \frac{2e}{h} \int_{E_F - e\frac{V}{2}}^{E_F + e\frac{V}{2}} dE\ T(E) \qquad (2)$$

Therefore the electrical conductance $G=I/V$ is obtained by averaging $T(E)$ over an energy window of width $eV$ centred upon the Fermi energy. The Fermi functions can then be Taylor expanded over the range $eV$ to give the electrical conductance in the zero-voltage but finite temperature limit by

$$G=I/V=G_0 \int_{-\infty}^{+\infty} dE\, T(E) \left(-\frac{df(E)}{dE}\right) \quad (3)$$

which represents a thermal average of $T(E)$ over an energy window of $k_B T$ where $k_B$ is the Boltzmann constant. As the normalised probability distribution $-\frac{df(E)}{dE}$ has a width of approximately $k_B T$ so in the limit of zero voltage and zero temperature $G=G_0 T(E_F)$. Fig. 3c (3d) shows the changes in the conductance (current) of the dumbell molecule placed between two gold electrodes for different angles between the rotor and molecular backbone. The conductance and therefore current is reduced significantly at $\theta_A = 60°$ and $\theta_B = 240°$ where the minima of $U_{BSC}$ occur.

Figure 4 The weighted current in the blue curve from Eqn.5 and the *NDR* (shown by the green curve as *dI/dV*) for applied bias between 0 and 1 *V*. The prominent *NDR* features are seen at a bias voltage between 0.0 and 0.1 volts and 0.6 and 0.7 volts for the device.

To include the effect of the parallel electric field due to the source-drain voltage *V* and induced rotation of the rotor, we use the energy lanscape $U_{BSC}(\theta,V)$ to construct the probability function

$$p(\theta,V)=A\, \exp\left(-\frac{U_{BSC}(\theta,V)}{k_B T}\right) \quad (4)$$

where *A* is a normalisation constant and compute the average current at voltage *V* using the relation

$$\langle I\rangle=\frac{2e}{h}\int_{-\infty}^{\infty} dE \int_{0}^{2\pi} d\theta\, p(\theta,V) T(E,V)\bigl(f_L(E)-f_R(E)\bigr) \quad (5)$$

The blue curve in Fig. 4 shows the weighted current at room temperature for applied biases between *-1* and *1 V*. By differentiating the current with respect to the bias voltage *V*, one obtains the differential conductance (green dashed line) of the device, which clearly shows regions of negative differential resistance (*NDR*) behaviour arising from the change in the energy landscape. The higher NDR effect occurs in the bias interval of [0.6, 0.7] *V*, although there is also a smaller NDR region at low bias voltage ~0.05 V.

At zero temperature, the rotation angle coincides with the global minimum of the energy curves. For example in figure 2a, at zero bias (black curve) the energy minimum corresponds to an angle of 240 degrees, whereas at for a field of 3.5*V/nm* parallel to the backbone, (blue curve) the energy minimum corresponds to 410 degrees. This demonstrates that such a field can cause the pendant group to rotate through 170 degrees. On the other hand, figure 2b shows that a field in the opposite direction does not shift the global minimum and therefore does not cause the pendant group to rotate. Similarly figure 2c shows that a field perpendicular to the backbone can shift the minimum to 360 degrees. It is these conformational change which cause the NDR, because the gating of the backbone due to the dipole moment of the pendant group is angle dependent. It should be noted that the applied field does not rotate the pendant group by 360 degrees. Nevertheless the voltage-dependent the energy landscapes $U_{BSC}(\theta,V)$ shown in figure 2 and the associated changes in the distribution of rotation angles $p(\theta,V)$ is sufficient to produce NDR. At finite temperature, the minima in $U_{BSC}(\theta,V)$ correspond to the most-probable rotation angles, but other angles can be sampled, according to the Boltzmann factor in equation (4).

**Conclusion**

We have examined the change in conformation of a molecular-scale rotator attached via a backbone to two $C_{60}$ anchor groups, which in turn are connected to gold electrodes. Our aim was to determine if an applied source-drain bias could cause the equilibrium angle of the rotator to change, leading to a non-linear current-voltage relation. Our results confirm that such a non-linearity indeed occurs and is strong enough to lead to a pronounced negative differential resistance region at relatively-low bias in the range [0.6, 0.7] volts. The underlying mechanism is that the dipole moment of the pendant group electrostatically gates the backbone states and this gating is angle dependent. Such NDR behaviour is potentially of interest for molecular-scale electronic applications such as single-molecule Gunn oscillators. The device studied in this paper utilises $C_{60}$ terminal groups attached to gold electrodes. These groups reduce the overall magnitude of the current and therefore for the future, it would be of interest to improve this proof of principle device, by utilising alternative combinations of terminal groups and electrodes, which increase the current. One such possibility would be planar anchor groups on graphene electrodes, which are currently under development in a number of groups [50-52] and allow the imposition of an external electric field via a nearby gate.

**Acknowledgment**

This work is supported by UK EPSRC grants EP/K001507/1, EP/J014753/1, EP/H035818/1 and the European Union Marie-Curie Network MOLESCO.

# References

1. Hiratsuka, Y.; T. Tada; K. Oiwa; T. Kanayama; T.Q. Uyeda, *Biophysical Journal* **2001**. *81*(3): p. 1555-1561. doi: 10.1016/S0006-3495(01)75809-2
2. Van den Heuvel, M.G.; C. Dekker, *Science* **2007**. *317*(5836): p. 333-336. doi: 10.1126/science.1139570
3. Fujimoto, K.; M. Kitamura; M. Yokokawa; I. Kanno; H. Kotera; R. Yokokawa, *Acs. Nano.* **2012**. *7*(1): p. 447-455. doi: 10.1021/nn3045038
4. Aoyama, S.; M. Shimoike; Y. Hiratsuka, *Proc. Natl. Acad. Sci.* **2013**. *110*(41): p. 16408-16413. doi: 10.1073/pnas.1306281110
5. Taira, S.; Y.Z. Du; Y. Hiratsuka; T.Q. Uyeda; N. Yumoto; M. Kodaka, *Biotechnology and bioengineering* **2008**. *99*(3): p. 734-739. doi: 10.1002/bit.21618
6. Song, W.; H. Möhwald; J. Li, *Biomaterials* **2010**. *31*(6): p. 1287-1292. doi: 10.1016/j.biomaterials.2009.10.026
7. Ionov, L.; M. Stamm; S. Diez, *Nano Lett.* **2005**. *5*(10): p. 1910-1914. doi: 10.1021/nl051235h
8. Van den Heuvel, M.G.; M.P. De Graaff; C. Dekker, *Science* **2006**. *312*(5775): p. 910-914. doi: 10.1126/science.1124258
9. Hess, H.; J. Clemmens; J. Howard; V. Vogel, *Nano Lett.* **2002**. *2*(2): p. 113-116. doi: 10.1021/nl015647b
10. Bachand, G.D.; H. Hess; B. Ratna; P. Satir; V. Vogel, *Lab Chip* **2009**. *9*(12): p. 1661-1666. doi: 10.1039/B821055A
11. Fischer, T.; A. Agarwal; H. Hess, *Nat. Nanotechnol.* **2009**. *4*(3): p. 162-166. doi: 10.1038/nnano.2008.393
12. Schliwa, M.; G. Woehlke, *Nature* **2003**. *422*(6933): p. 759-765. doi: 10.1038/nature01601
13. Bruns, C.J.; J.F. Stoddart, *Acc. Chem. Res.* **2014**. *47*(7): p. 2186-2199. doi: 10.1021/ar500138u
14. Kumar, K.R.S.; T. Kamei; T. Fukaminato; N. Tamaoki, *Acs. Nano.* **2014**. *8*(5): p. 4157-4165. doi: 10.1021/nn5010342
15. Gao, W.; J. Wang, *Nanoscale* **2014**. *6*(18): p. 10486-10494. doi: 10.1039/C4NR03124E
16. Browne, W.R.; B.L. Feringa, *Nat. Nanotechnol.* **2006**. *1*(1): p. 25-35. doi: 10.1038/nnano.2006.45
17. Balzani, V.; A. Credi; M. Venturi, *Molecular devices and machines: concepts and perspectives for the nanoworld.* **2008**: John Wiley & Sons.
18. Garcia-Garibay, M.A., *Proc. Natl. Acad. Sci. U.S.A.* **2005**. *102*(31): p. 10771-10776. doi: 10.1073/pnas.0502816102
19. Karlen, S.D.; M.A. Garcia-Garibay, *Amphidynamic crystals: Structural blueprints for molecular machines*, in *Molecular Machines*. 2005, Springer. p. 179-227.
20. Davis, A.P., *Angewandte Chemie International Edition* **1998**. *37*(7): p. 909-910. doi: 10.1002/(SICI)1521-3773(19980420)37:7<909::AID-ANIE909>3.0.CO;2-X
21. Kelly, T.R.; J.P. Sestelo; I. Tellitu, *The Journal of Organic Chemistry* **1998**. *63*(11): p. 3655-3665. doi: 10.1021/jo9723218
22. Hugel, T.; N.B. Holland; A. Cattani; L. Moroder; M. Seitz; H.E. Gaub, *Science* **2002**. *296*(5570): p. 1103-1106. doi: 10.1126/science.1069856
23. Liu, Y.; A.H. Flood; P.A. Bonvallet; S.A. Vignon; B.H. Northrop; H.-R. Tseng; J.O. Jeppesen; T.J. Huang; B. Brough; M. Baller, *J. Am. Chem. Soc.* **2005**. *127*(27): p. 9745-9759. doi: 10.1021/ja051088p
24. Coskun, A.; M. Banaszak; R.D. Astumian; J.F. Stoddart; B.A. Grzybowski, *Chemical Society Reviews* **2012**. *41*(1): p. 19-30. doi: 10.1039/C1CS15262A
25. Bruns, C.J.; J.F. Stoddart, *Nat. Nanotechnol.* **2013**. *8*(1): p. 9-10. doi: 10.1038/nnano.2012.239
26. Morin, J.-F.; Y. Shirai; J.M. Tour, *Organic letters* **2006**. *8*(8): p. 1713-1716. doi: 10.1021/ol060445d
27. Cumings, J.; A. Zettl, *Science* **2000**. *289*(5479): p. 602-604. doi: 10.1126/science.289.5479.602
28. Yu, M.-F.; B.I. Yakobson; R.S. Ruoff, *The Journal of Physical Chemistry B* **2000**. *104*(37): p. 8764-8767. doi: 10.1021/jp002828d
29. Kolmogorov, A.N.; V.H. Crespi, *Phys. Rev. Lett.* **2000**. *85*(22): p. 4727-4730. doi: 10.1103/PhysRevLett.85.4727
30. Zheng, Q.; Q. Jiang, *Phys. Rev. Lett.* **2002**. *88*(4): p. 045503. doi: 10.1103/PhysRevLett.90.055504
31. Forró, L., *Science* **2000**. *289*(5479): p. 560-561. doi: 10.1126/science.289.5479.560
32. Williams, P.A.; S.J. Papadakis; A.M. Patel; M.R. Falvo; S. Washburn; R. Superfine, *Phys. Rev. Lett.* **2002**. *89*(25): p. 255502. doi: 10.1103/PhysRevLett.89.255502
33. Rivera, J.L.; C. McCabe; P.T. Cummings, *Nano Lett.* **2003**. *3*(8): p. 1001-1005. doi: 10.1021/nl034171o
34. Sangtarash, S.; H. Sadeghi; M.T. Ahmadi; M.H. Ghadiry; S. Anwar; R. Ismail, *J. Comput. Theor. Nanosci.* **2012**. *9*(10): p. 1554-1557. doi: 10.1166/jctn.2012.2243
35. Grace, I.M.; S.W. Bailey; C.J. Lambert, *Phys. Rev. B.* **2004**. *70*(15): p. 153405. doi: 10.1103/PhysRevB.70.153405
36. Bailey, S.W.D.; I. Amanatidis; C.J. Lambert, *Phys. Rev. Lett.* **2008**. *100*(25): p. 256802. doi: 10.1103/PhysRevLett.100.256802
37. Cai, K.; H. Yin; Q.H. Qin; Y. Li, *Nano Lett.* **2014**. *14*(5): p. 2558-2562. doi: 10.1021/nl5003608
38. Fennimore, A.M.; T.D. Yuzvinsky; W.-Q. Han; M.S. Fuhrer; J. Cumings; A. Zettl, *Nature* **2003**. *424*(6947): p. 408-410. doi: 10.1038/nature01823

39. Bourlon, B.; D.C. Glattli; C. Miko; L. Forró; A. Bachtold, *Nano Lett.* **2004**. *4*(4): p. 709-712. doi: 10.1021/nl035217g
40. Niguès, A.; A. Siria; P. Vincent; P. Poncharal; L. Bocquet, *Nat. Mater.* **2014**. *13*(7): p. 688-693. doi: 10.1038/nmat3985
41. Leary, E.; M.T. González; C. van der Pol; M.R. Bryce; S. Filippone; N. Martín; G. Rubio-Bollinger; N. Agraït, *Nano Lett.* **2011**. *11*(6): p. 2236-2241. doi: 10.1021/nl200294s
42. Soler, J.M.; E. Artacho; J.D. Gale; A. García; J. Junquera; P. Ordejón; D. Sánchez-Portal, *J. Phys.: Condens. Mat.* **2002**. *14*(11): p. 2745. doi: 10.1088/0953-8984/14/11/302
43. Dion, M.; H. Rydberg; E. Schröder; D.C. Langreth; B.I. Lundqvist, *Phys. Rev. Lett.* **2004**. *92*(24): p. 246401. doi: 10.1103/PhysRevLett.92.246401
44. Langreth, D.C.; M. Dion; H. Rydberg; E. Schröder; P. Hyldgaard; B.I. Lundqvist, *Int. J. Quantum Chem.* **2005**. *101*(5): p. 599-610. doi: 10.1002/qua.20315
45. Landauer, R., *Zeitschrift für Physik B Condensed Matter* **1987**. *68*(2-3): p. 217-228. doi: 10.1007/BF01304229
46. Ferrer, J.; C. Lambert; V. García-Suárez; D.Z. Manrique; D. Visontai; L. Oroszlany; R. Rodríguez-Ferradás; I. Grace; S. Bailey; K. Gillemot; H. Sadeghi; L.A. Algharagholy, *New J Phys* **2014**. *16*(9): p. 093029. doi: 10.1088/1367-2630/16/9/093029
47. Sadeghi, H.; J.A. Mol; C.S. Lau; G.A.D. Briggs; J. Warner; C.J. Lambert, *Proc. Natl. Acad. Sci.* **2015**. *112*(9): p. 2658–2663. doi: 10.1073/pnas.1418632112
48. Lambert, C., *Chemical Society Reviews* **2015**(44): p. 875-888. doi: 10.1039/c4cs00203b
49. Santiago, M.; M. David Zsolt; M.G.-S. Víctor; H. Wolfgang; J.H. Simon; J.L. Colin; J.N. Richard, *Nanotechnology.* **2009**. *20*(12): p. 125203. doi: 10.1088/0957-4484/20/12/125203
50. Sadeghi, H.; S. Sangtarash; C.J. Lambert, *Beilstein Journal of Nanotechnology* **2015**. *6*: p. 1413. doi: 10.3762/bjnano.6.146
51. Mol, J.A.; C.S. Lau; W.J. Lewis; H. Sadeghi; C. Roche; A. Cnossen; J.H. Warner; C.J. Lambert; H.L. Anderson; G.A.D. Briggs, *Nanoscale* **2015**. *7*(31): p. 13181-13185. doi: 10.1039/C5NR03294F
52. Prins, F.; A. Barreiro; J.W. Ruitenberg; J.S. Seldenthuis; N. Aliaga-Alcalde; L.M.K. Vandersypen; H.S.J. van der Zant, *Nano Lett.* **2011**. *11*(11): p. 4607-4611. doi: 10.1021/nl202065x